\documentclass{epl2}
\usepackage[latin1]{inputenc}
\usepackage{setspace}
\usepackage{amsmath}
\usepackage{amsfonts}
\usepackage{amssymb}
\usepackage{graphicx}
\usepackage{wrapfig}
\usepackage{subfig}

\title{Chiral d-wave superconductivity in the heavy-fermion compound $\rm CeIrIn_5$}
\author{K. Maki \and A. Raghavan \and S. Haas}
\institute{Department of Physics \& Astronomy, University of Southern California, Los Angeles, CA 90089-0484}
\abstract{Recent thermal conductivity measurements in the heavy-fermion compound
$\rm CeIrIn_5$ indicate that its superconducting order parameter 
is very different from $\rm CeCoIn_5$.  Here we show that these 
experiments are consistent 
with chiral d-wave symmetry, i.e. $\Delta(\vec{k})\sim e^{\pm i\phi}\cos(ck_z)$.}

\pacs{74.70.Tx}{Heavy-fermion superconductors}
\pacs{74.25.Fy}{Transport properties}
\pacs{71.27.+a}{Strongly correlated electron systems; heavy fermions}

\begin{document}
\maketitle

The discovery of antiparamagnon mediated superconductivity in the 115 heavy-fermion compounds 
$\rm CeTIn_5$, where T represents Co, Ir, Rh, or a mixture of these, 
has recently opened up a new avenue to unconventional nodal superconductivity\cite{petrovic}.  These 
strongly interacting materials are 
characterized by a plethora of competing ground states in addition to superconductivity, including conventional 
and unconventional spin density wave (SDW) phases\cite{knebel}. Among the 115 compounds, the currently most well 
studied is $\rm CeCoIn_5$ for which a d-wave superconducting order parameter
$\Delta(\vec{k})\sim\cos(2\phi)=\hat{k}_x^2-\hat{k_y^2}$ has been identified\cite{izawa,aoki,won,miclea}.  
Indeed, there are many parallels between $\rm CeCoIn_5$ and the high-$T_c$ cuprates, including
(a) a layered quasi-two-dimensional Fermi surfaces\cite{maehira},
(b) d-wave superconductivity, and
(c) d-wave spin density wave order in the pseudogap phase\cite{hu,maki,won2}.

Recent thermal conductivity measurements \cite{shakeripour,ohira} indicating an order parameter symmetry
in $\rm CeIrIn_5$ very different from the one in $\rm CeCoIn_5$ came as a big surprise.  
An initial analysis of this data suggested a hybrid $E_g$ gap, 
$\Delta(\vec{k})\sim Y_{2,\pm 1}(\theta,\phi)$, 
based on the assumption that the Fermi surface is three-dimensional. However, the Fermi surface of $\rm CeIrIn_5$ 
is in fact quasi-two-dimensional, as known from band structure analysis\cite{maehira,ohira}.  Therefore, one needs
to consider instead superconductivity in
layered structures, similar as discussed in Refs. \cite{won3,won4}. In this case, only 
$f=e^{\pm i\phi}\sin(\chi)$ (chiral d-wave) with $\chi = ck_z$ or $f\sim \sin(\chi)$ (non-chiral p-wave) are
consistent with the observed thermal conductivity data \cite{shakeripour}.  
The magnitudes of the d-wave and chiral d-wave/non-chiral p-wave
order parameters $|\Delta(\vec{k})|$ are shown in Fig. 1.

\begin{figure}
  \centering
    \includegraphics[width=3cm]{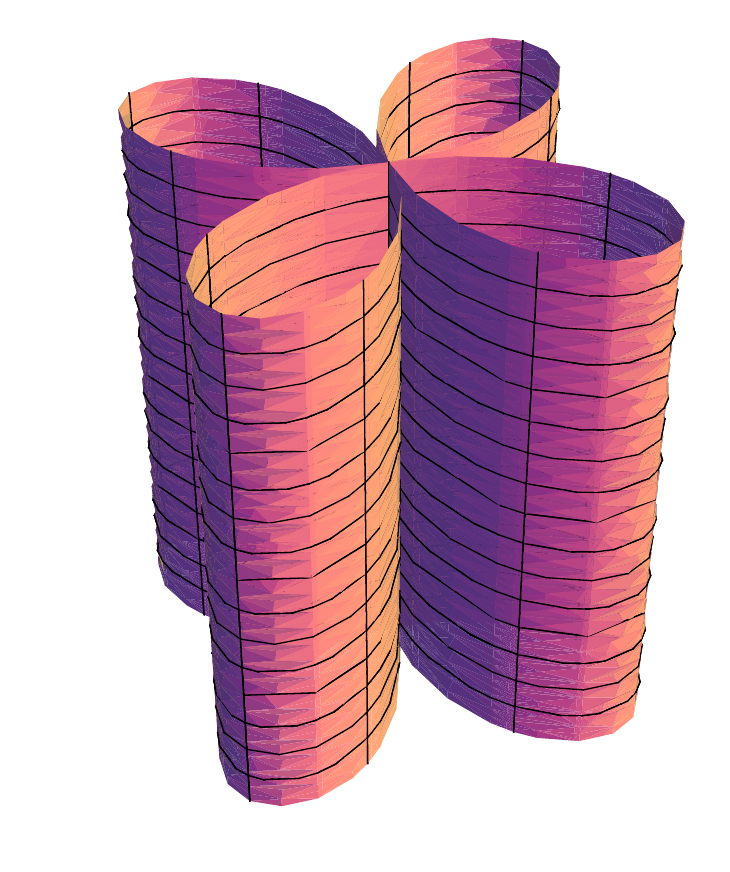}
    \includegraphics[width=2.5cm]{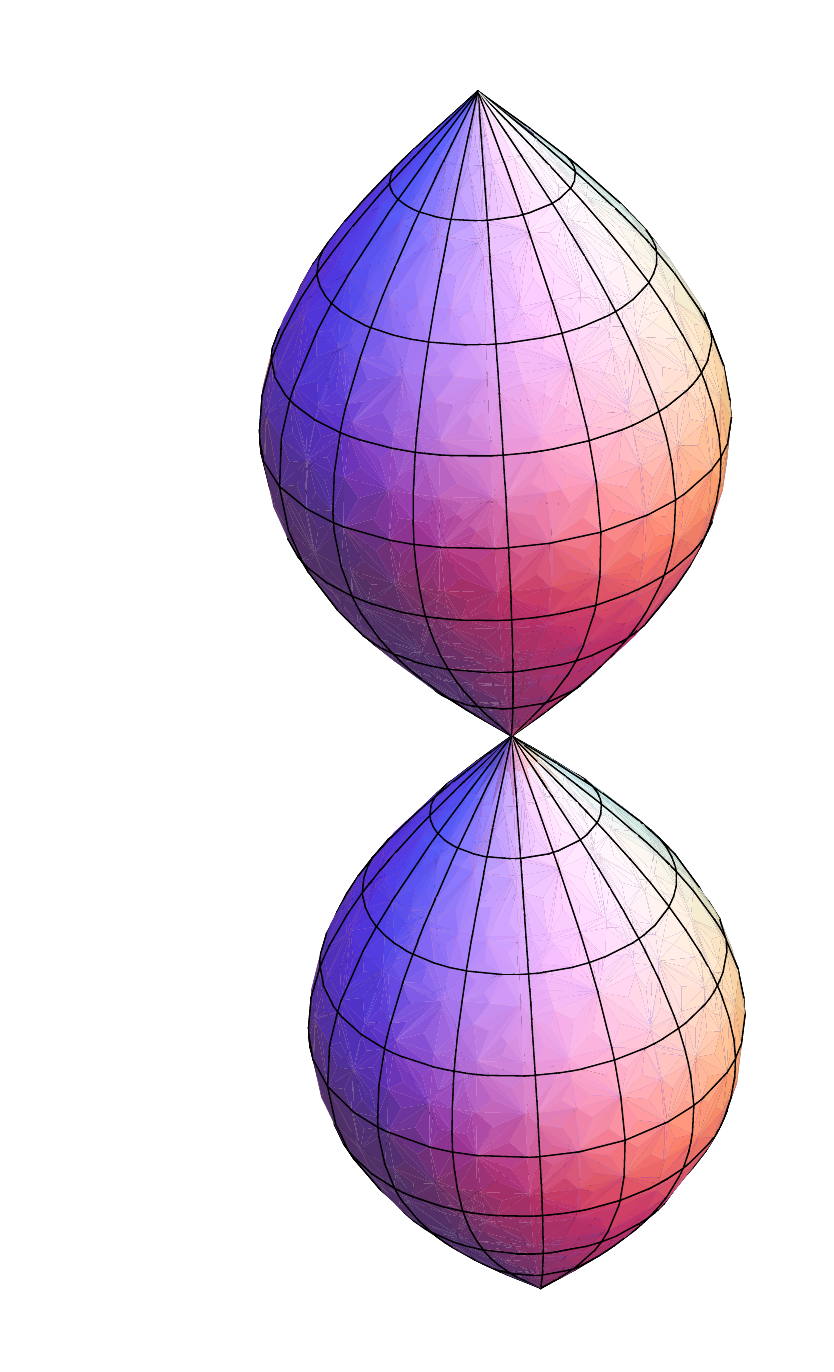}
  \caption{Magnitude of order parameters $|\Delta(\vec{k})|$ for d-wave (left) 
and chiral d-wave/non-chiral p-wave (right) superconductors.}
  \label{fig:delta_k}
\end{figure}

In the following, we present a theoretical analysis based on a generalized BCS model
that properly accounts for a quasi-two-dimensional Fermi surface and a chiral d-wave superconducting order
parameter.
The thermal conductivity is computed following the receipe given in Refs. \cite{won4,yang}.  
Here, we assume for simplicity that the quasiparticle scattering is due to impurities.  
Furthermore, we consider the physically relevant limit $\Gamma/\Delta\ll 1$, where $\Gamma$ 
is the quasiparticle scattering rate in the normal state and $\Delta (=0.856K)$ is the maximum 
value of the energy gap at $T=0K$.  This $\Delta$ is the weak-coupling value for 
nodal superconductors\cite{won3,wonmaki}.

Let us begin by considering the 
zero-temperature limit. For quasi-two-dimensional structures, the thermal
conductivity
strongly depends on the direction within the material.
Therefore we will discuss the cases
$\vec{q}\|\vec{a}$ (in-plane) and $\vec{q}\|\vec{c}$
(out-of-plane) separately.
Here $\vec{q}$ denotes the heat current.
For $\vec{q}\|\vec{a}$, one obtains 
\begin{equation}
\dfrac{\kappa^a}{\kappa^a_n}=\dfrac{2\Gamma_a}{\pi\Delta}
\end{equation}
and similarly for $\vec{q}\|\vec{c}$
\begin{equation}
\dfrac{\kappa^c}{\kappa^c_n}=2\left(\dfrac{\Gamma_c}{\Delta}\right)^2,
\end{equation}
where $\Gamma_a$ and $\Gamma_c$ denote the in-plane and out-of-plane 
scattering rates respectively. 
Eq.1 describes the universal heat conduction as discovered by 
P. Lee\cite{lee,sun}, whereas Eq.2 is very different.  The strength 
of the impurity scattering can be extracted directly from the 
experimental data show in Fig. 2 of Ref.\cite{shakeripour}, from 
which we can deduce that $\dfrac{\Gamma}{\Delta}=0.19635$.  Furthermore,
from the observed anisotropy of the thermal conductivity, we can infer
the ratio of the Fermi velocities along the c-axis and the a-b plane,
i.e. $\dfrac{v_c}{v_a}=0.66$, which is very similar to $\dfrac{v_c}{v_a}=0.5$ 
extracted for $\rm CeCoIn_5$\cite{hu}.  Then, for $T\neq 0K$ but 
$\dfrac{T}{\Delta}\ll 1$, we obtain in the regime $T\gg\Gamma$,
\begin{equation}
\dfrac{\kappa^a(T)}{\kappa^a_n(T)}=\dfrac{27}{2\pi^2}\zeta(3)\left(\dfrac{T}{\Delta}\right)+O\left(\dfrac{T}{\Delta}\right)^3,
\end{equation}
and
\begin{equation}
\dfrac{\kappa^c(T)}{\kappa^c_n(T)} = \dfrac{45^2}{4\pi^2}\zeta(5)\left(\dfrac{T}{\Delta}\right)^3+O\left(\dfrac{T}{\Delta}\right)^5.
\end{equation}
This is consistent with the experimental observation of a dominant in-plane 
heat conductivity propertional to the temperature, and a subdominant 
out-of-plane conductivity. 

In order to connect these finite-temperature results with the above
equations for $T=0$, we use an interpolation formula which applies 
in the regime for $T/\Delta(T)\ll 1$. The resulting low-temperature 
thermal conductivities are then given by
\begin{equation}
\dfrac{\kappa^a(T)}{\kappa^a_n(T)} =\dfrac{2\Gamma_a}{\pi\Delta}\left(1+\left(\dfrac{27}{4\pi}\zeta(3)\dfrac{T}{\Gamma_a}\right)^2\right)^{1/2}
\end{equation}
and
\begin{equation}
\dfrac{\kappa^c(T)}{\kappa^c_n(T)} = 2\left(\dfrac{\Gamma_c}{\Delta}\right)^2\left(1+\left(\dfrac{45^2}{8\pi^2}\zeta(5)\left(\dfrac{T}{\Gamma_c}\right)^2\left(\dfrac{T}{\Delta}\right)\right)^2\right)^{1/2}
\end{equation}
respectively.

In Fig. 2, we compare these dependencies with the experimental data 
reported in Ref.\cite{shakeripour}.  A fit of the low-temperature regimes
yields good agreement with $\dfrac{\Gamma_c}{\Gamma_a}=0.5592$.  
Evidently, the quasi-particle scattering rate is somewhat anisotropic in 
the present system.  Here, the temperature dependence of the gap function
$\Delta (T)$, is approximated by 
\begin{equation}
\Delta(T)=2.14T_c\left[1-\left(\dfrac{T}{T_c}\right)^3\right]^{1/2}
\end{equation}
with $T_c=0.4K$, which is known to be a very good approximation 
for d-wave superconductors\cite{dora}.

\begin{figure}
  \centering
    \includegraphics[width=6.0cm]{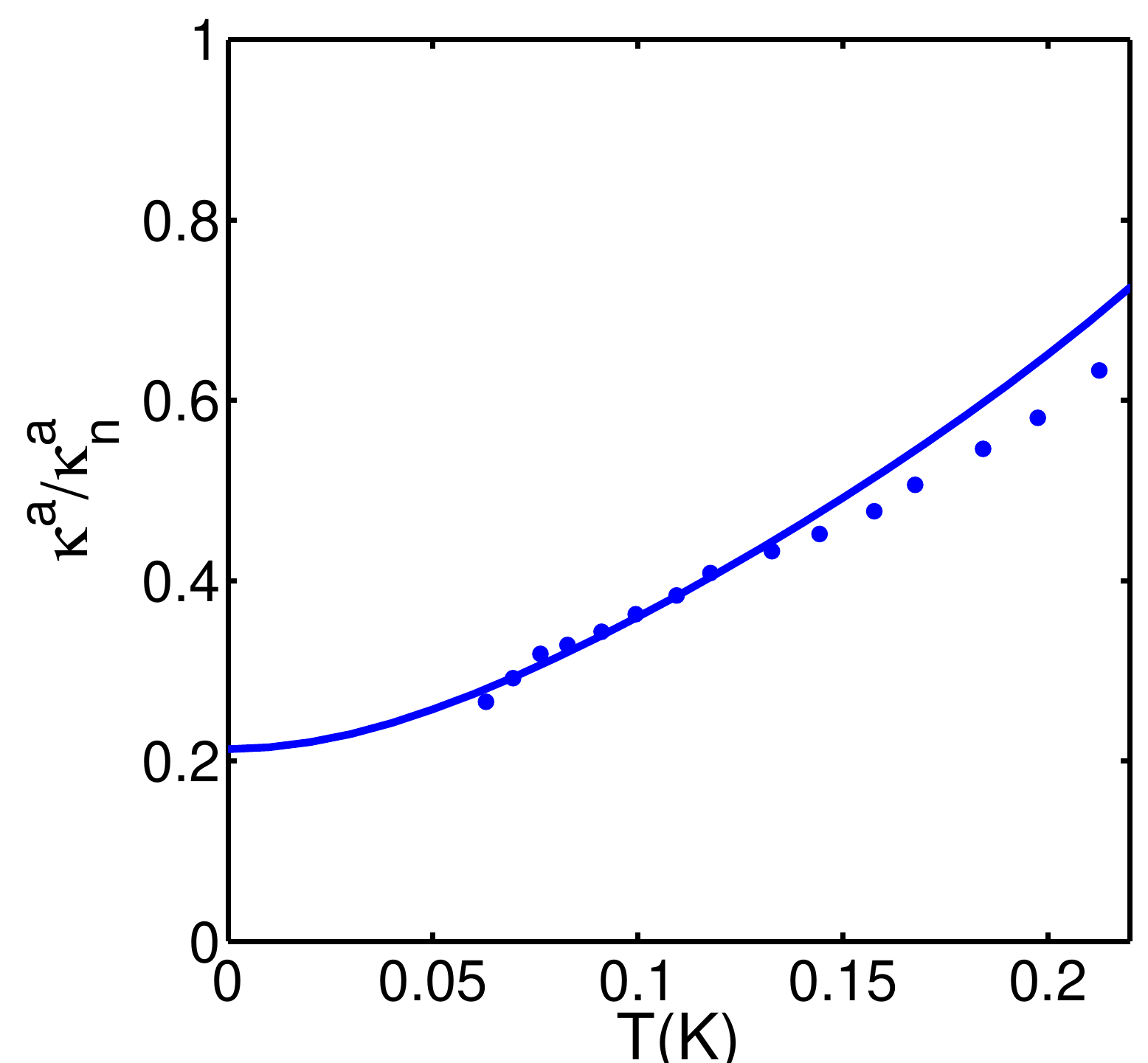}
    \includegraphics[width=6.0cm]{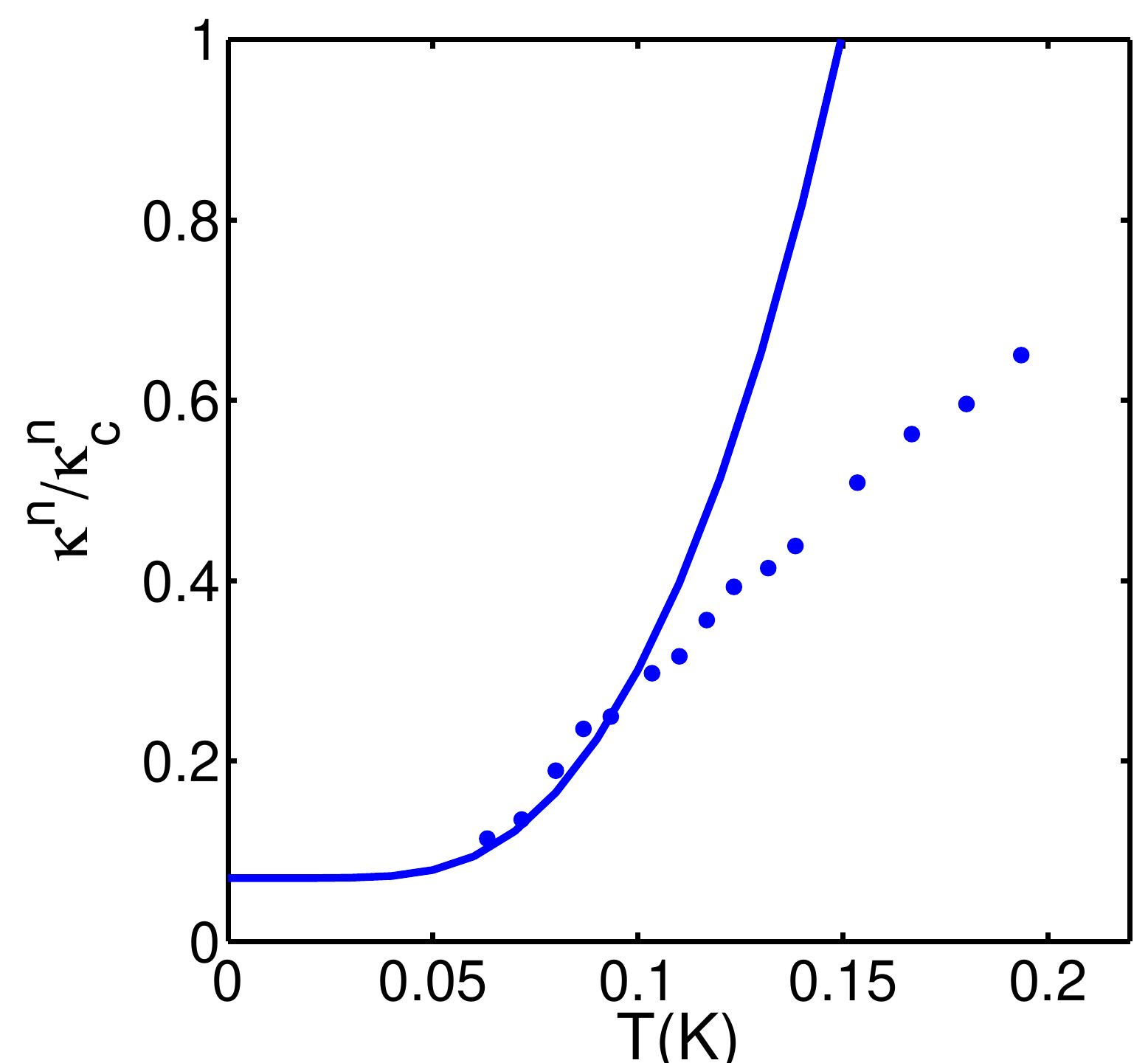}
  \caption{Thermal conductivity in the $\vec{q}\| \vec{a}$ (in-plane) and 
$\vec{q}\| \vec{c}$ (out-of-plane) directions.  $T_c=0.4 K$.  
The symbols represent experimental data from \cite{shakeripour}, and the 
solid lines are low-temperature fits using Eqs. 5 and 6.}
  \label{fig:two}
\end{figure}

Similarly, the ratio $\kappa^c(T)/\kappa^a(T)$ can be computed and 
compared to the experiments. Within our model, it is given by
\begin{equation}
\kappa^c(T)/\kappa^a(T)=0.2703\left[\dfrac{1+\left(\dfrac{45^2}{8\pi^2}\zeta(5)\right)^2\dfrac{T^6}{\Gamma_c^4\Delta^2}}
{1+\left(\dfrac{27}{4\pi}\zeta(3)\right)^2\left(\dfrac{T}{\Gamma_a}\right)^2}\right]^{1/2}
\end{equation}
which is shown in Fig. 3 along with the thermal conductivity measurements of Ref. \cite{shakeripour}.
\begin{figure}
  \centering
    \includegraphics[width=8cm]{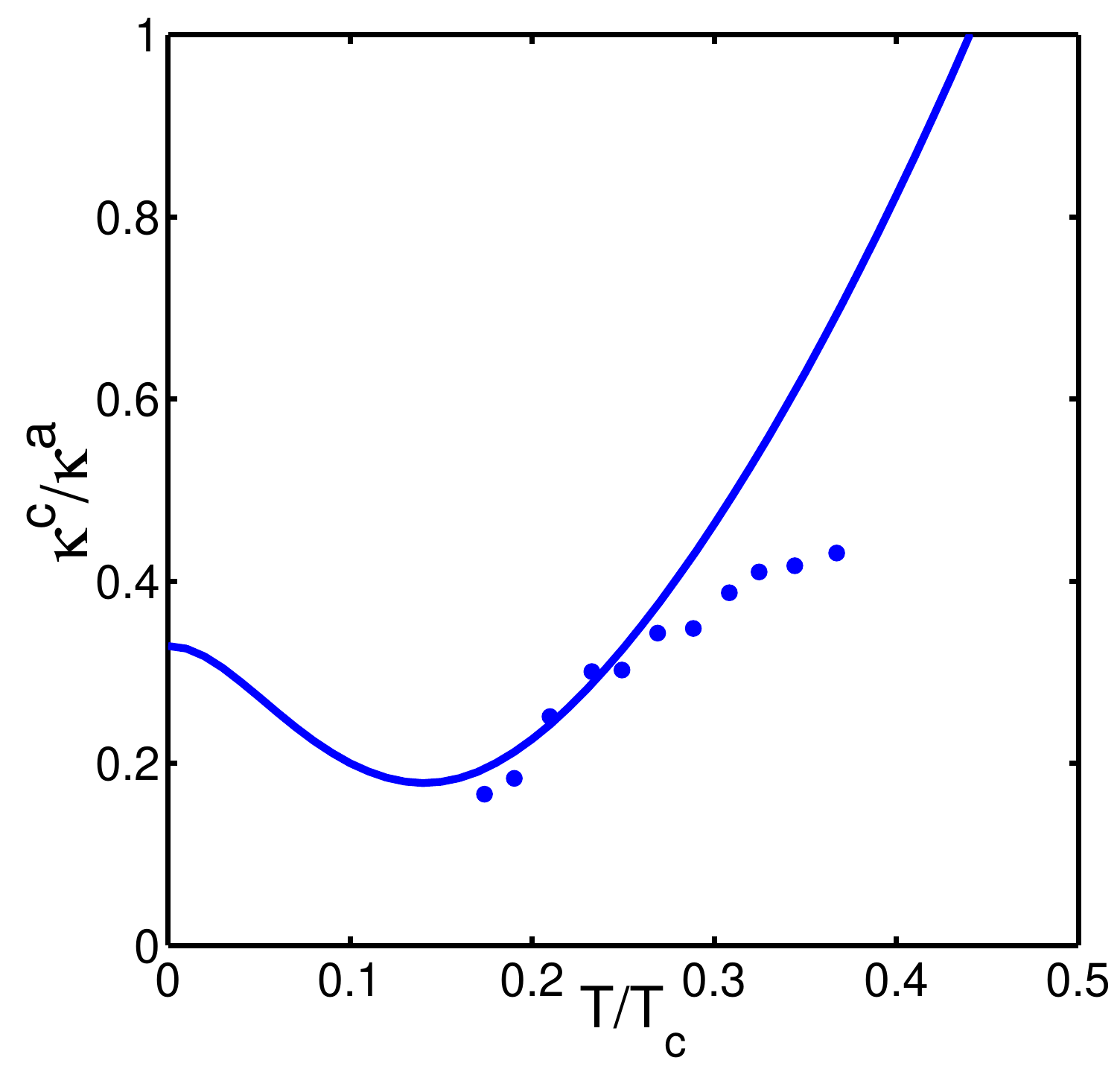}
  \caption{Ratio of thermal conductivities of the $\vec{q}\| c$ 
and $\vec{q}\|\vec{a}$ direction, plotted as a function of $T/T_c$.  
The symbols represent the experimental data from Ref.\cite{shakeripour}.}
  \label{fig:three}
\end{figure}

These expressions give a very reasonable description of the thermal 
conductivity for $T/T_c\leq 0.3$.  We note that a similarly good description 
of the thermal conductivity is given by the hybrid gap proposed in Ref.
\cite{shakeripour}.  At higher temperatures, $T/T_c\geq 0.3$,
our simple model fails to 
describe the measured thermal conductivity, possibly due to the fact that 
phonons begin to play an important role as we approach $T\rightarrow T_c$.  
Nevertheless, we can conclude that chiral d-wave SC is consistent with 
the experimental data of Refs. \cite{shakeripour,ohira} in the relevant 
low-temperature regime. Note also, that our calculations predict an 
interesting upturn in the ratio $\kappa^c(T)/\kappa^a(T)$ as the temperature
is further lowered. This prediction can be scrutinized experimentally, and 
may serve as a means to distinguish the present theory from the hybrid
gap model that was proposed earlier.

In the present context, the unconventional superconducting order
 in $\rm CeRhIn_5$ is of
great interest. Let us briefly contemplate on the doped
case.   Inspecting Fig. 3 of Ohira-Kawamura et al\cite{ohira} 
we may conclude that the order parameter in $\rm CeRh_{1-x}Co_xIn_5$ 
should be d-wave SC with an
angular dependence $f = \cos(2\phi)$, whereas the order 
parameter in $CeRh_{1-x}Ir_xIn_5$ is consistent with
chiral d-wave superconductivity with an angular dependence
$f = e^{\pm i\phi}\cos(\chi)$.  Therefore, the above approach will provide a 
basis to identify the many competing 
phases of the 115 compounds.  Also, the 
phase diagrams for $\rm CeRh_{1-x}Co_xIn_5$ and $\rm CeRh_{1-x}Ir_xIn_5$ 
in Ref.\cite{ohira} are of great interest for the perspective of the Gossamer
superconductivity, i.e. a phase with competing order parameters\cite{won2,won5,maki2}.
We observe that 
(a) the incommensurate phases in both $\rm CeRh_{1-x}Co_xIn_5$ and 
$\rm CeRh_{1-x}Ir_xIn_5$ are conventional spin-density waves, 
(b) the commensurate phase in $\rm CeRh_{1-x}Co_xIn_5$ and
the incommensurate+commensurate phase in $\rm CeRh_{1-x}Ir_xIn_5$ 
have d-wave symmetry.
Therefore, there is a wide region where d-wave superconductivity 
coexists with unconventional nodal spin density wave order. 

In summary, we have successfully applied a nodal weak-coupling BCS theory
to fit recent experimental 
data on the directional thermal conductivity of $\rm CeRhIn_5$. We 
find that in contrast to $\rm CeCoIn_5$, which has plain d-wave order,
this compound is consistent with chiral 
d-wave superconductivity. Furthermore, this technique will allow us
to identify the many different phases which were recently discovered 
in doped derivatives of these materials. 

\acknowledgements
We would like to thank Y. Matsuda for very informative discussions 
on the superconductivity in $\rm CeIrIn_5$.

\end{document}